\documentclass{elsart}
\usepackage{cancel}
\usepackage{txfonts}
\usepackage{psfrag}
\usepackage{amsfonts}
\usepackage{lscape}
\usepackage{url}
\usepackage[dvips]{graphicx}

\begin{document}
\begin{frontmatter}

\title{Stability in generic mitochondrial models}

\author{Pete Donnell\thanksref{cor1}\thanksref{ref1}} 
\author{Murad Banaji\thanksref{cor1}\thanksref{ref1}}
\author{Stephen Baigent\thanksref{ref2}}
\address[ref1]{Department of Medical Physics and Bioengineering, University College London, Gower Street, London WC1E 6BT}
\address[ref2]{Department of Mathematics, University College London, Gower Street, London WC1E 6BT}
\thanks[cor1]{Funded by an EPSRC/MRC grant to the MIAS IRC (Grant Ref: GR/N14248/01)}

\begin{abstract}
In this paper, we use a variety of mathematical techniques to explore existence, local stability, and global stability of equilibria in abstract models of mitochondrial metabolism. The class of models constructed is defined by the biological description of the system, with minimal mathematical assumptions. The key features are an electron transport chain coupled to a process of charge translocation across a membrane. In the absence of charge translocation these models have previously been shown to behave in a very simple manner with a single, globally stable equilibrium. We show that with charge translocation the conclusion about a unique equilibrium remains true, but local and global stability do not necessarily follow. In sufficiently low dimensions -- i.e. for short electron transport chains -- it is possible to make claims about local and global stability of the equilibrium. On the other hand, for longer chains, these general claims are no longer valid. Some particular conditions which ensure stability of the equilibrium for chains of arbitrary length are presented. 

\end{abstract}


\end{frontmatter}

\section{Introduction}

The processes of electron transport and oxidative phosphorylation in mitochondria are of vital biological importance, being central to cellular respiration and hence energy production in most eukaryotic cells. Summaries of these processes can be found in many modern biochemistry textbooks such as \cite{GandG} or \cite{bhagavan}. The basic features of mitochondrial electron transport and oxidative phosphorylation are now well understood, but elucidation of many of the detailed mechanisms is still in progress \cite{belevich}.

Mitochondrial electron transport occurs via a series of coupled redox reactions in the mitochondrial inner membrane. After the initial reduction of a first electron donor (e.g. $\mathrm{NADH}$ or $\mathrm{FADH_2}$ produced by glycolysis and the TCA cycle) electrons are transferred from substrate to substrate, finally being accepted by oxygen. During some of these electron transfers a second process takes place -- protons are pumped across the mitochondrial inner membrane producing a proton gradient across this membrane. These protons then return down their gradient, either passively (termed leak current) or through a particular enzyme, $\mathrm{ATP}$ synthase, leading to the phosphorylation of ADP.

Generic models of electron transport chains were explored in \cite{banajiJTB}, where the main emphasis was on the input-output response of such models. In the simplest case, where the proton gradient across the membrane was ignored, these models were found to have very simple behaviour -- at all physically meaningful parameter values there was a single, globally stable, equilibrium. In \cite{banajimathchem}, this result was shown to generalise to the case of electron transfer networks with more general topology than a chain. On the other hand in the more biologically realistic case -- where the build up of a proton gradient has an inhibitory effect on electron transport -- analysis of the models proved harder. In this paper we analyse in more detail the behaviour in this case.

Before discussing generic models, it is worth mentioning that there are several detailed models of electron transport and oxidative phosphorylation such as \cite{korzeniewski2}, \cite{korzeniewski1}, \cite{farmery}, \cite{beard}. These ordinary differential equation models have been designed with numerical data in mind, and reflecting the complexity of the processes involved, the functional forms are quite involved. Our interest in mitochondria was originally inspired by analysis and simulation of some of these numerical models, but the approach here is quite different, and more akin to work in \cite{banajiJTB}, \cite{banajimathchem}, \cite{leenheer}. The generic model we construct could be instantiated in a great variety of numerical models, and the claims we make are valid for all possible instances of the generic model.

\section{The model}

\subsection{The basic reaction scheme}
The basic reaction scheme of interest here was described in some detail in \cite{banajiJTB} but will be summarised here. Assume that there are $n$ substrates, each of which can exist in an oxidised state $\mathrm{A_i}$ and a reduced state $\mathrm{B_i}$ so that 
\[
\mathrm{A_i} + e^{-} \leftrightharpoons \mathrm{B_i}
\]
Further, assume that protons can exist in two compartments -- the mitochondrial matrix (where they are termed $\mathrm{H}^{+}_{m}$), and the intermembrane space (where they are termed $\mathrm{H}^{+}_{e}$) -- with the possibility of transfers of the form
\[
\mathrm{H}^{+}_{m} \leftrightharpoons \mathrm{H}^{+}_{e}
\]
We are interested in reactions which are in general the combination of three processes, a reduction, an oxidation, and the transport of some protons across the membrane. So for example, if substrate $\mathrm{A_i}$ is reduced to $\mathrm{B_i}$, $\mathrm{B_j}$ is oxidised to $\mathrm{A_j}$, and $p$ protons are pumped across the mitochondrial membrane we get the half reactions
\[
\mathrm{A_i} + e^{-} \leftrightharpoons \mathrm{B_i}, \quad \mathrm{B_j} \leftrightharpoons \mathrm{A_j} + e^{-}  \quad\mbox{and}\quad  p\mathrm{H}^{+}_{m} \leftrightharpoons p\mathrm{H}^{+}_{e} 
\]
which combine to give
\[
\mathrm{A_i} + \mathrm{B_j} + p\mathrm{H}^{+}_{m} \leftrightharpoons \mathrm{A_j} + \mathrm{B_i} + p\mathrm{H}^{+}_{e}
\]
We also allow the possibility that a reducing/oxidising agent may be external to the model giving reactions such as
\[
\mathrm{A_i} + p\mathrm{H}^{+}_{m} \leftrightharpoons \mathrm{B_i} + p\mathrm{H}^{+}_{e}
\quad \mbox{or}\quad
\mathrm{B_i} + p\mathrm{H}^{+}_{m} \leftrightharpoons \mathrm{A_i} + p\mathrm{H}^{+}_{e}
\]

A set of reactions of the kind just described can be combined into a network of reactions. A chain structure (as opposed to a more general network) derives from the assumption that each oxidised substrate accepts an electron from only one donor, and each reduced substrate transfers its electron to only one acceptor. This introduces a natural ordering on the substrates, so that for $i < n$, the $i$th substrate is able to donate electrons to the $(i+1)$th substrate, while for $i > 1$, the $i$th substrate is able to accept electrons from the $(i-1)$th substrate. The first substrate is able to accept electrons from outside the chain (reflecting the initial reduction of $\mathrm{NADH}$ or $\mathrm{FADH_2}$), and the $n$th substrate is able to donate electrons to an acceptor outside the chain (reflecting the action of $\mathrm{O}_2$). 

Thus there are $n+1$ redox reactions and the $i$th reaction has forward rate $f_i$. We make no assumptions about the sign of the $f_i$, potentially allowing reactions to be reversible. For $i\leq n$, the $i$th reaction involves the reduction $\mathrm{A_i}$, and for $i \geq 2$, the $i$th reaction involves the oxidation of $\mathrm{B_{i-1}}$. We define $p_i$ as the number of protons pumped across the mitochondrial membrane by the $i$th reaction. Assuming that the quantities $p_i$ are constant discounts the possibility of ``redox slip'' \cite{brand}, which does not appear to be very important in normal circumstances \cite{canton}. A quantity $\psi$ can be defined so that transfer of a single proton across the membrane creates one unit of $\psi$. $\psi$ can take any real value and is a strictly increasing function of the electrical/chemical gradient against which protons are pumped across the membrane, generally termed the proton motive force. 

Finally, reflecting the combined effect of proton leak and ADP phosphorylation, there is a process with rate $L$ representing the ``decay'' of $\psi$. When there is no gradient, no protons leak through the membrane, so that $L(0) = 0$. Further $L$ is assumed to be strictly increasing in $\psi$. 

\begin{figure}[ht]
\centering
\begin{minipage}{\textwidth}
\begin{center}
\includegraphics[width=0.8\textwidth]{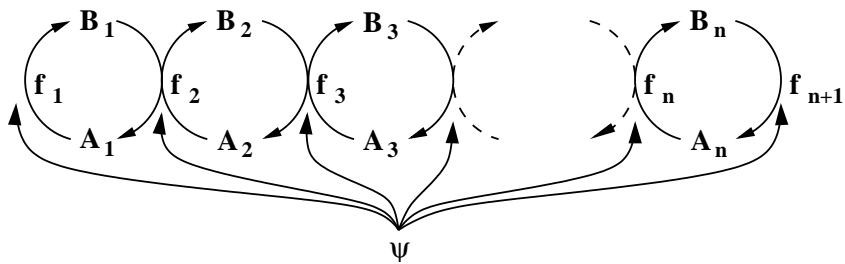}
\end{center}
\end{minipage}
\begin{minipage}{\textwidth}
\caption{\label{basicscheme}A schematic representation of the reaction network. The quantities $\mathrm{A_i}$ and $\mathrm{B_i}$ refer to oxidised and reduced states of the substrates. The functions $f_i$ define the forward rates of reaction of the $n+1$ coupled redox reactions. The quantity $\psi$ represents the electrical and chemical gradient across the mitochondrial membrane, which has an inhibitory effect on any redox reactions which involve proton pumping.}
\end{minipage}

\end{figure}

The structure of the model is illustrated in Figure~\ref{basicscheme}.

Because the total quantity -- oxidised plus reduced -- of any substrate in the chain is conserved, reduced forms of the substrates are not explicitly introduced. Instead, the concentration of $\mathrm{A_i}$ is referred to as $x_i$, and the total concentration of $\mathrm{A_i} + \mathrm{B_i}$ is assumed constant at $m_i$. We arrive at a model of the form:

\begin{equation}
\left. \begin{array}{rcl}
\dot{x_1} & = & -f_1(x_1, \psi) + f_2(x_1, x_2, \psi) \\
\dot{x_i} & = & -f_i(x_{i-1}, x_i, \psi) + f_{i+1}(x_i, x_{i+1}, \psi) \quad i= 2,\ldots,n-1 \\
\dot{x_n} & = & -f_n(x_{n-1}, x_n, \psi) + f_{n+1}(x_n, \psi) \\
\dot{\psi} & = & \sum \limits_{i=1}^{n+1} p_i f_i - L(\psi)
\end{array}\ \ \right\} \label{DEp}
\end{equation}

The phase space of this system is defined by the equations:
\begin{eqnarray*}
0 \leq & x_i & \leq m_i \quad i = 1, \ldots, n \\
-\infty < & \psi & < \infty
\end{eqnarray*}
and is hence $n+1$ dimensional, being the product of a closed $n$-dimensional box and the real line.

\subsection{Assumptions}
\label{secassumptions}

All the functions $f_i$, along with $L$, are assumed to be \(C^1\) (once differentiable in all their arguments with continuous derivatives). The following notation is used for the derivatives of the functions $f_i$:
\begin{equation}
f_{ij} \equiv \frac{\partial f_{i}}{\partial x_j}\,, \quad F_{ij} \equiv -f_{ij}\,, \quad f_{i\psi} \equiv \frac{\partial f_{i}}{\partial \psi}\,, \quad F_{i\psi} \equiv -f_{i\psi} \label{function_labels}
\end{equation}
At finite substrate concentrations, all reaction rates are finite, so that at any fixed $\psi$ each $f_i$ is bounded on its domain of definition.

Since $\psi$ represents a potential against which some of the reactions must do work, the following relations are obtained:

\begin{equation}
\label{neg_deriv2}
f_{i\psi} < 0 \mbox{ if } p_i \not = 0 \quad \mbox{and} \quad f_{i\psi} = 0 \mbox{ if } p_i = 0
\end{equation}
If $p_i \not = 0$, then $\psi$ inhibits the forward reaction and we assume that sufficiently large values of $\psi$ make the reaction rate arbitrarily small or negative, i.e.
\begin{eqnarray*}
\lim_{\psi\to\infty} f_i(\cdot, \psi) \leq 0 \qquad i = 1, n+1\\
\lim_{\psi\to\infty} f_i(\cdot, \cdot, \psi) \leq 0 \qquad i = 2, \ldots, n
\end{eqnarray*}
This reflects the fact that the energy required to pump a proton against a chemical and electrical gradient becomes large as the gradient increases. Similarly $-\psi$ inhibits the backward reaction so that:
\begin{eqnarray*}
\lim_{\psi\to-\infty} f_i(\cdot, \psi) \geq 0 \qquad i = 1, n+1\\
\lim_{\psi\to-\infty} f_i(\cdot, \cdot, \psi) \geq 0 \qquad i = 2, \ldots, n
\end{eqnarray*}

The following equations imply that no reaction can proceed in the absence of any of its substrates:
\begin{equation}
\left. \begin{array}{rcl}
f_1(0, \cdot) & = & 0  \\
f_i(\cdot, 0, \cdot) & = & 0  \quad i = 2, \cdots, n \\
f_i(m_{i-1}, \cdot, \cdot) & = & 0 \quad i = 2, \cdots, n\\
f_{n+1}(m_n, \cdot) & = & 0
\end{array}\ \ \right\}
\end{equation}

The final set of conditions imply that increased substrate concentration increases the rate of reaction unless one of the substrates is entirely absent:
\begin{equation}
\left. \begin{array}{rcl}
f_{11} & > & 0 \\
f_{ii} & \geq & 0 \mbox{ and } f_{ii} > 0 \mbox{ if }x_{i-1} < m_{i-1} \quad i = 2, \cdots, n \\
f_{i+1, i} & \leq & 0  \mbox{ and } f_{i+1, i} > 0 \mbox{ if }x_{i+1} > 0 \quad i = 1, \cdots, n-1 \\
f_{n+1, n} & < & 0
\end{array}\ \ \right\} \label{condlast}
\end{equation}

The fact that the first and final inequalities are always strict implies that there is always some electron donor to reduce the initial substrate, and some electron acceptor to oxidise the final substrate, and ensures nondegenerate behaviour. The assumptions from (\ref{condlast}) mean that $f_{ii}$, $F_{ij}$ and $F_{i\psi}$ as defined in (\ref{function_labels}) are all nonnegative. The definition of these nonnegative quantities is solely to simplify later arguments. 

\section{General behaviour of the system}

In this section we outline some properties of the model that hold regardless of the number \(n\) of redox pairs.

\subsection{Boundedness of solutions}

It is convenient to define an $n \times (n+1)$ matrix which can be regarded as a stoichiometric matrix for the redox reactions:
\[
S \equiv
\left[
\begin{array}{ccccc}
-1 & 1 & \cdots & 0&0\\
0 & -1 & \cdots & 0&0\\
\vdots&\vdots&\ddots&\vdots&\vdots\\
0&0&\cdots&-1&\,\,1
\end{array}
\right]
\]

Defining the vector of reactant concentrations $\mathbf{x} = [x_1, x_2, \ldots, x_n]^T$, the vector of reaction rates $\mathbf{v}(\mathbf{x}, \psi) = [f_1, f_2, \ldots f_{n+1}]^T$, and the nonnegative vector $P \equiv [p_1, \ldots, p_{n+1}]^T$, we can rewrite the system of equations (\ref{DEp}) more briefly as
\begin{eqnarray*}
\label{vec}
\dot {\mathbf x} & = & S\mathbf{v}({\mathbf x}, \psi)\\
\dot{\psi} & = & P^T\mathbf{v}({\mathbf x}, \psi) - L(\psi)
\end{eqnarray*}

We now show that all forward trajectories of the system are bounded. Since the phase space is bounded in $\mathbf{x}$, what needs to be shown is that all trajectories enter a bounded region in the $\psi$ direction. This amounts to showing that $\dot \psi > 0$ for $\psi$ sufficiently large and negative, and that $\dot\psi<0$ for $\psi$ sufficiently large and positive. By assumption, for any given $i$, either $p_i = 0$ or $f_{i\psi}$ is strictly negative and $\lim_{\psi\to\infty} f_i(\cdot, \cdot, \psi) \leq 0$, $\lim_{\psi\to-\infty} f_i(\cdot, \cdot, \psi) \geq 0$. This in turn implies that \(\lim_{\psi \rightarrow \infty} P^T\mathbf{v}({\mathbf x}, \psi) \leq 0\) and \(\lim_{\psi \rightarrow \infty} P^T\mathbf{v}({\mathbf x}, \psi) \geq 0\). In addition $L$ is strictly increasing from zero as $\psi$ increases. Thus for any fixed value of $\mathbf{x}$, \(\lim_{\psi \rightarrow \infty}P^T\mathbf{v}({\mathbf x}, \psi) - L(\psi) < 0\) and \(\lim_{\psi \rightarrow -\infty}P^T\mathbf{v}({\mathbf x}, \psi) - L(\psi) > 0\). Define $\psi_{0}(\mathbf{x})$ as the value of $\psi$ at which $P^T\mathbf{v}({\mathbf x}, \psi) - L(\psi) = 0$. $\psi_{0}(\mathbf{x})$ is uniquely defined since $P^T\mathbf{v}({\mathbf x}, \psi) - L(\psi)$ is strictly decreasing. By the implicit function theorem, $\psi_{0}(\mathbf{x})$ is a differentiable function since  $P^T\mathbf{v}({\mathbf x}, \psi) - L(\psi)$ is a differentiable function of $\mathbf{x}$. Since it has a compact domain, $\psi_{0}(\mathbf{x})$ achieves a maximum value which we call $\psi_{max}$, and a minimum value which we call $\psi_{min}$. By these definitions, $\dot\psi(\psi, {\mathbf x}) < 0$ for all $\psi > \psi_{max}$, and $\dot\psi(\psi, {\mathbf x}) > 0$ for all $\psi < \psi_{min}$.

Thus all trajectories enter a closed box, $\mathcal{B}$, bounded by the hyperplanes $x_i = 0$, $x_i = m_i$, $\psi = \psi_{min}$ and $\psi = \psi_{max}$, and this box forms a trapping region for the system in all dimensions.

\subsection{The Jacobian}
\label{secjacobian}

Direct calculation gives that the Jacobian, $J$, of the system is:

\vspace{0.5cm}
\[
J = \left[
\begin{array}{ccccc}
-f_{11}-F_{21} & f_{22} & \cdots & 0 & F_{1\psi} - F_{2\psi}\\
F_{21} & -f_{22}-F_{32} & \cdots & 0 & F_{2\psi}- F_{3\psi}\\
\vdots & \vdots & \ddots & \vdots & \vdots \\
0 & 0 & \cdots & -f_{nn}-F_{n+1,n} & F_{n\psi} - F_{n+1,\psi}\\
p_1f_{11}\!-\!p_2F_{21}\, & \,p_2f_{22}\!-\!p_3F_{32}\, & \,\cdots\, & \,p_{n}f_{nn}\!-\!p_{n+1}F_{n+1,n}\, & \,-L_{\psi}\!-\!\sum\limits_{i=1}^{n+1}p_iF_{i\psi}\\
\end{array}
\right]
\]

Here $L_\psi \equiv \frac{\mathrm{d} L}{\mathrm{d} \psi}$. The structure of this Jacobian can be made clearer by defining two further quantities: A nonnegative vector in $\mathbb{R}^n$,  $F \equiv [F_{1\psi}, \ldots, F_{n\psi}]^T$; and an $(n+1) \times n$ matrix
\[
V \equiv \frac{\partial \mathbf{v}}{\partial \mathbf{x}} =
\left[
\begin{array}{ccccc}
f_{11} & 0 & 0 & \cdots &0\\
-F_{21} & f_{22} & 0 & \cdots &0\\
0 & -F_{32} & f_{33} & \cdots &0\\
\vdots&\vdots&\vdots&\ddots&\vdots\\
0&0&0&\cdots&f_{nn}\\
0&0&0&\cdots&-F_{n+1,n}
\end{array}
\right]
\]
Then the Jacobian can be written in the block form:
\begin{equation}
\label{Jeq1}
J = \left[
\begin{array}{cc}
SV & SF\\
P^TV & -P^TF - L_\psi \\
\end{array}
\right]
\end{equation}

$SV$ is the Jacobian of the system without feedback, which is tridiagonal, and can easily be shown to have real negative eigenvalues \cite{banajiJTB}. It was shown in \cite{banajiSIAM} that the structures of $S$ and $V$ along with the nonnegativity of $P$ and $F$ imply that $J$ is a so called $P^{(-)}$ matrix (see Appendix~\ref{appdefs} for the definition)\footnote{The nondegeneracy conditions presented in \cite{banajiSIAM} are met because the $n$th substrate is terminal, and all substrates are able to transfer electrons along the chain to the $n$th substrate.}. This result is independent of $n$, the length of the chain. It has the consequence that the system is injective; this is discussed further in the next section.

The fact that $J$ is a $P^{(-)}$ matrix has another consequence of importance to us: It means that its eigenvalues are excluded from a certain wedge around the positive real axis: If $\lambda = re^{i\theta}$ is an eigenvalue of an $m \times m$ $P$ matrix, then it is proved in \cite{kellogg} that:
\[
|\theta - \pi| > \pi/m
\]
and equivalently for a  $P^{(-)}$ matrix,
\[
|\theta| > \pi/m
\]
Clearly when $m=2$, this means that both eigenvalues lie in the left half plane, so that $2 \times 2$ $P^{(-)}$ matrices are Hurwitz stable (see Appendix~\ref{appdefs} for a definition of ``Hurwitz stable'' which we will abbreviate to ``Hurwitz''). However for $m > 2$, $P^{(-)}$ matrices may be unstable.

\subsection{A unique equilibrium}

The existence of a unique equilibrium for this system was shown in \cite{banajiJTB} by a direct method. It also follows from the arguments presented above: That an equilibrium must exist follows, by the Brouwer fixed point theorem, from the existence of the compact, convex, trapping region, $\mathcal{B}$ constructed above; That this equilibrium must be unique follows from the fact that the Jacobian is a $P^{(-)}$ matrix, and hence the system is injective \cite{gale}. Thus as our first result we can state that

\begin{quote}
Electron transport chains coupled to charge translocation across a membrane have exactly one equilibrium.
\end{quote}

It is interesting that the possibility of multistability is immediately ruled out. However this in itself does not tell us whether all trajectories must necessarily converge to the unique equilibrium, or whether periodic or chaotic behaviour is still possible.

\section{Stability of the equilibrium}

In this section, we analyse stability of the unique equilibrium, starting with low dimensions (i.e. short chains). For two dimensions we prove that the equilibrium is globally asymptotically stable. In three dimensions we show that the addition of an extra, reasonable, constraint implies that the equilibrium is locally stable, and further constraints ensure that it is globally stable. We then demonstrate that these constraints do not suffice to guarantee stability in four dimensions and higher. Finally, we outline some additional special conditions that guarantee the Jacobian is Hurwitz in all dimensions.

\subsection{The system in two dimensions}

The system in 2D consists of a single redox pair subject to a reduction process and an oxidation process, both possibly coupled to proton translocation across the membrane. It takes the form

\begin{eqnarray*}
\dot{x_1} & = & -f_1(x_1, \psi) + f_2(x_1, \psi) \\
\dot{\psi} & = & p_1f_1 + p_2f_2 - L(\psi)
\end{eqnarray*}

The Jacobian of the system in this case is:

\vspace{0.5cm}
\begin{equation}
J_2 = \left[
\begin{array}{cc}
-f_{11}-F_{21} & F_{1\psi} - F_{2\psi}\\
p_1f_{11}\!-\!p_2F_{21}\, & \,-L_{\psi}\!-\!p_1F_{1\psi}-\!p_2F_{2\psi}\\
\end{array}
\right]
\end{equation}

We have already mentioned that 2D $P^{(-)}$ matrices are Hurwitz stable, and it follows that the matrices $J_2$ are Hurwitz stable (This can also be shown with a direct calculation).

Since $J_2$ is Hurwitz stable {\em everywhere}, not just at the unique equilibrium, the Markus-Yamabe Theorem (e.g. \cite{fessler}, \cite{glutsyuk}, \cite{gutierrez}) ensures that the equilibrium is globally stable. We also offer an alternative, elementary, proof of global stability. By the Poincar\'{e}-Bendixson Theorem (see, for example, \cite{ciesielski}), $\omega$-limit sets of a flow on compact subsets of $\mathbb{R}^2$ must either contain equilibria or consist of a periodic orbit. In this case we can rule out the possibility of periodic orbits: The divergence of the vector field is equal to
\[
Tr(J) = -f_{11}-F_{21}-p_1F_{1\psi}-p_2F_{2\psi}-L_\psi
\]
which is negative. Thus the vector field satisfies the Dulac criterion (e.g. \cite{guckenheimer}) and there are no periodic orbits. We know that there is only one equilibrium, which is locally stable, and therefore there are no heteroclinic or homoclinic orbits either. Since every forward trajectory enters the box $\mathcal{B}$, the unique equilibrium must be the $\omega$-limit of every trajectory, and is hence globally stable. 

\subsection{The system in three dimensions}
\label{3dstability}

Slightly more complex than the two dimensional system is the system in three dimensions which takes the form

\begin{eqnarray*}
\dot{x_1} & = & -f_1(x_1, \psi) + f_2(x_1, x_2, \psi) \\
\dot{x_2} & = & -f_2(x_1, x_2, \psi) + f_3(x_2, \psi) \\
\dot{\psi} & = & p_1f_1 + p_2f_2 + p_3f_3 - L(\psi)
\end{eqnarray*}

with Jacobian
\vspace{0.5cm}
\begin{equation}
\label{J3eq}
J_3 = \left[
\begin{array}{ccc}
-f_{11}-F_{21} & f_{22} & F_{1\psi} - F_{2\psi}\\
F_{21} & -f_{22}-F_{32} & F_{2\psi}- F_{3\psi}\\
p_1f_{11}\!-\!p_2F_{21}\, & \,p_2f_{22}\!-\!p_3F_{32}\, & \,-L_{\psi}\!-\!p_1F_{1\psi}-\!p_2F_{2\psi}-\!p_3F_{3\psi}\\
\end{array}
\right]
\end{equation}

As it stands, \(J_3\) is not always Hurwitz. For example, the Jacobian constructed using the following values: \(p_1=3, p_2=0, p_3=88, F_{1\psi}=33, F_{2\psi}=4, F_{3\psi}=0.6, f_{11}=23, f_{22}=3, F_{21}=94, F_{32}=76, L_{\psi}=6\) has two eigenvalues with positive real part.

$J_3$ can be shown to be Hurwitz everywhere in 3D provided one extra condition is met: $p_1$ and $p_3$ must have the same ordering as $F_{1\psi}$ and $F_{3\psi}$. For a real number \(z\), define the function
\begin{equation}
\mathrm{sign}(z) \equiv \left\{
\begin{array}{ll}
1 & (z > 0) \\ 0 & (z = 0) \\ -1 & (z < 0) 
\end{array}
\right.
\end{equation}

Then the ordering assumption translates to the following statement:
\begin{equation}
\mathrm{sign}(F_{3\psi} - F_{1\psi}) = \mathrm{sign}(p_3 - p_1) \label{hurwitz_sign_assumption}
\end{equation}

With this assumption, the Jacobian is everywhere Hurwitz, and hence the equilibrium is locally asymptotically stable. The proof is simple but requires some lengthy evaluations, and the details are presented in Appendix~\ref{local3D}.

Unlike in the 2D case it does not follow that the equilibrium is globally stable, since the Markus-Yamabe conjecture does not hold in dimensions greater than 2 \cite{cima}. However we can prove global stability in this case too subject to a strengthened version of the ordering assumption on the quantities $p_i$ and $F_{i\psi}$. We now require
\begin{equation}\
\label{orderass}
\mathrm{sign}(F_{i\psi} - F_{j\psi}) = \mathrm{sign}(p_i - p_j)
\end{equation}
for $i, j \in \{1, 2, 3\}$.

With this assumption we are able to use a version of Li and Muldowney's autonomous convergence theorem (Theorem 4.1 in \cite{muldowney}) to show that the unique equilibrium is globally stable. In order to use this theorem two concepts are needed:
\begin{enumerate}
\item The {\bf second additive compound} of a matrix
\item {\bf Logarithmic norms} of a matrix
\end{enumerate}
Both quantities are defined for square matrices. The second additive compound matrix of any $n \times n$ matrix $J$ is a square matrix of dimension $^nC_2$ which we will term $J^{[2]}$. Logarithmic norms are scalar quantities, and corresponding to any given matrix norm, there is a logarithmic norm. Unlike matrix norms, however, logarithmic norms may take negative values. The definitions are given in Appendix~\ref{appdefs}.

Consider a dynamical system with Jacobian $J(x)$ at some point of phase space $x$. Define \(\mathbf{J}\) to be the set of all these Jacobians. For our purposes, the autonomous convergence theorem states the following: If a logarithmic norm $\mu$ can be found such that
\begin{equation}
\mu(J^{[2]}) < 0 \mbox{ for all } J \in \mathbf{J} \label{autonomous_convergence_theorem}
\end{equation}
then the limit set of each bounded semi-trajectory of the dynamical system is an equilibrium.

Since all trajectories enter the trapping region $\mathcal{B}$ in our system, and since $\mathcal{B}$ contains a unique equilibrium, finding a suitable logarithmic norm satisfying (\ref{autonomous_convergence_theorem}) will suffice to prove global stability of the equilibrium.

The second additive compound in this case is:
\[J_3^{[2]} = \left[\begin{array}{ccc}
-f_{11}\!-\!F_{21}\!-\!f_{22}\!-\!F_{32} & F_{2\psi}- F_{3\psi} & -(F_{1\psi} - F_{2\psi}) \\
\,p_2f_{22}\!-\!p_3F_{32}\, & -f_{11}\!-\!F_{21}\!-\!L_{\psi}\!-\!\sum \limits_{i+1}^3p_iF_{i\psi} & f_{22} \\
-(p_1f_{11}\!-\!p_2F_{21}) & F_{21} & -f_{22}\!-\!F_{32}\!-\!L_{\psi}\!-\!\sum \limits_{i+1}^3p_iF_{i\psi}
\end{array}\right]\]

We will construct a logarithmic norm $\mu_T$ such that $\mu_T\left(J_3^{[2]}\right) < 0$. For a real $n \times n$ matrix, the logarithmic norm corresponding the usual $\|\cdot\|_1$ norm takes the form:
\[\mu_{1} = \max\limits_{i \in \{1, \ldots, n\}} \left(x_{ii} + \sum \limits_{k,k \neq i}|x_{ki}|\right)\]
From the definition it is clear that a matrix has negative logarithmic norm $\mu_1$ if and only if every diagonal entry is negative and it is strictly diagonally dominant in every column. Next we define a constant diagonal coordinate transformation
\[T = \left(\begin{array}{ccc}
1 & 0 & 0 \\
0 & \frac{1}{p_{max}} & 0 \\
0 & 0 & \frac{1}{p_{max}}
\end{array}\right) \]
where $p_{max} = \max\limits_{i \in \{1, 2, 3\}} (p_i)$. 

According to Lemma 2.2 of \cite{li_wang}, given any invertible transformation $T$, $\mu_T(M) \equiv \mu_1(T M T^{-1})$ defines a new logarithmic norm. In this case, since \(T\) is a diagonal matrix, the diagonal entries of \(M\) are the same as those of \(TMT^{-1}\). Thus in order to prove that $\mu_T(J_3^{[2]})< 0$, we need to show that $J' \equiv TJ_3^{[2]}T^{-1}$ is strictly diagonally dominant in every column.

For the first column, we have

\begin{eqnarray}
J'_{11} + \left|J'_{21}\right| + \left|J'_{31}\right| & = & -f_{22} -F_{32} -f_{11} -F_{21} \nonumber \\
& & \hspace{1cm} + \left|\frac{p_2}{p_{max}}f_{22} -\frac{p_3}{p_{max}}F_{32}\right| +\left|\frac{p_2}{p_{max}}F_{21} -\frac{p_1}{p_{max}}f_{11}\right| \nonumber
\end{eqnarray}
It can easily be seen that the term on the right hand side is negative since for any two nonnegative scalars $|a-b| \leq \max\{|a|, |b|\}$.

For the second column, we have
\[J'_{22} + \left|J'_{12}\right| + \left|J'_{32}\right| = -\sum_{i=1}^3p_iF_{i\psi}-L_\psi -f_{11} + p_{max}\left|F_{2\psi} -F_{3\psi}\right| \]

For the final column, we have
\[J'_{33} + \left|J'_{13}\right| + \left|J'_{23}\right| = -\sum_{i=1}^3p_iF_{i\psi} -L_\psi -F_{32} +  p_{max}\left|F_{2\psi} -F_{1\psi}\right|\]
In order to show that the right hand sides of the last two expressions are negative we need to show in each case that our ordering assumption (\ref{orderass}) implies that the final term (which may be positive) is dominated in magnitude by the other terms.

Note that $|F_{i\psi}-F_{j\psi}| \leq \max\{F_{i\psi}, F_{j\psi}\} \leq \max\limits_{k \in \{1, 2, 3\}}(F_{k\psi})$. Then there are only three cases:
\begin{enumerate}
\item if $p_{max} = p_1$, then $p_{max}\left|F_{2\psi} -F_{3\psi}\right| \leq p_1F_{1\psi}$, and  $p_{max}\left|F_{2\psi} -F_{1\psi}\right| \leq p_1F_{1\psi}$.
\item if $p_{max} = p_2$, then $p_{max}\left|F_{2\psi} -F_{3\psi}\right| \leq p_2F_{2\psi}$, and  $p_{max}\left|F_{2\psi} -F_{1\psi}\right| \leq p_2F_{2\psi}$.
\item if $p_{max} = p_3$, then $p_{max}\left|F_{2\psi} -F_{3\psi}\right| \leq p_3F_{3\psi}$, and  $p_{max}\left|F_{2\psi} -F_{1\psi}\right| \leq p_3F_{3\psi}$.
\end{enumerate}

Each of these possibilities leads to the same conclusion -- that $J'_{ii} + \sum \limits_{k,k \neq i}|J'_{ki}| < 0$ for each $i$. Hence we have $\mu_T\left(J_3^{[2]}\right) < 0$.

This result means that if the ordering assumption (\ref{orderass}) holds, then the unique equilibrium is globally stable. The ordering assumption itself has the following reasonable physical meaning which we would expect to be fulfilled in practice: If redox reaction $i$ is involved in pumping more protons across the membrane than redox reaction $j$, then reaction $i$ is correspondingly more inhibited by $\psi$ than reaction $j$. It is interesting to note however that this assumption is not necessary to prove global stability in the 2D case. It is also unknown to us whether the weaker assumption (\ref{hurwitz_sign_assumption}), which guarantees that the Jacobian is everywhere Hurwitz, actually guarantees global stability in 3D. 

\subsection{Unstable examples in higher dimensions}
The ordering assumption (\ref{orderass}) does not guarantee global or even local stability of the equilibrium in dimensions greater than 3. It is easy to construct counterexamples. For example, in four dimensions, the Jacobian constructed by choosing $p_1 = 2$, $p_2 = p_3 = 0$, $p_4 = 73$, $F_{1\psi} = 167$, $F_{2\psi} = F_{3\psi} = 0$, $F_{4\psi} = 176$, $f_{11} = 4$, $f_{22} = 7$, $f_{33} = 1$, $F_{21} = 32$, $F_{32} = 64$, $F_{43} = 174$, $L_\psi = 33$, satisfies all the constraints, including the ordering assumption on the values of $p_i$ and $F_{i\psi}$. However it has, two eigenvalues with positive real part.

We make the following remarks:
\begin{enumerate}
\item By continuity, the fact that a non-Hurwitz Jacobian can be constructed in 4 dimensions guarantees that such examples also exist in all higher dimensions.
\item Systems with non-Hurwitz Jacobian satisfying the ordering assumption (\ref{orderass}) seem to be rare. Through use of an automated computer script running in the open source numerical computation program Scilab \cite{scilab}, counterexamples in dimension 4 were found by randomly choosing values for the different terms in the Jacobian, such that all the assumptions were satisfied. Out of hundreds of millions of sets of values, less than ten were non-Hurwitz.
\item The counterexamples found appear always to be close to breaking the ordering assumption. For instance, in the example shown, $p_4$ is much greater than $p_1$, whereas $F_{4\psi}$ is close in magnitude to $F_{1\psi}$.
\end{enumerate}

\subsection{A special case: Reaction rates dependent on potentials}

In this section we consider an interesting assumption which ensures that the Jacobian is Hurwitz everywhere (and hence the unique equilibrium is locally stable). The assumption is as follows: 
\begin{enumerate}
\item Associated with each half reaction is some ``potential'': In the case of a redox reaction of the form $\mathrm{A_i} + e^{-} \leftrightharpoons \mathrm{B_i}$, a potential means any strictly increasing scalar function of $[\mathrm{A_i}]$; In the case of a charge transfer across a membrane a potential means any strictly increasing scalar function of $\psi$.
\item The rate of any full reaction depends only on the {\bf sum} of the potentials for the half reactions involved, and is a strictly decreasing function of this sum.
\end{enumerate}

This assumption can be interpreted, loosely, as saying that the energetics of the system determine the reaction rates.  For example, consider the electron transfer coupled to some proton pumping
\[
\mathrm{A_i} + \mathrm{B_j} + p\mathrm{H}^{+}_{m} \leftrightharpoons \mathrm{A_j} + \mathrm{B_i} + p\mathrm{H}^{+}_{e}
\]
derived from the half reactions
\[
\mathrm{A_i} + e^{-} \leftrightharpoons \mathrm{B_i}, \quad \mathrm{B_j} \leftrightharpoons \mathrm{A_j} + e^{-}  \quad\mbox{and}\quad  p\mathrm{H}^{+}_{m} \leftrightharpoons p\mathrm{H}^{+}_{e} 
\]
In this case, the assumption would imply that the forward rate of the combined reaction can be written $f(-g_j(x_j) + g_i(x_i) - pg_\psi(\psi))$ where the only stipulation is that $f$, $g_i$, $g_j$ and $g_\psi$ are strictly increasing in their arguments. When this assumption is made about all reaction rates in the system, the full system becomes:

\begin{eqnarray*}
\label{DEp1red}
\dot{x_1} & = & -f_1(g_1(x_1)- p_1g_\psi(\psi)) + f_2(-g_1(x_1)+ g_2(x_2)- p_2g_\psi(\psi)) \\
\label{DEpired}
\dot{x_i} & = & -f_i(-g_{i-1}(x_{i-1})+ g_i(x_i)- p_ig_\psi(\psi)) +\nonumber\\
&& \hspace{2cm} f_{i+1}(-g_i(x_i)+ g_{i+1}(x_{i+1})- p_{i+1}g_\psi(\psi)) \quad i = 2, \ldots, n\\
\label{DEpnred}
\dot{x_n} & = & -f_n(-g_{n-1}(x_{n-1})+ g_n(x_n)- p_ng_\psi(\psi)) + f_{n+1}(-g_n(x_n) - p_{n+1}g_\psi(\psi))\\
\label{DEpn1red}
\dot{\psi} & = & \sum_{i=1}^{n+1}p_if_i - L(\psi)
\end{eqnarray*}
The term $f_i(-g_{i-1}(x_{i-1})+ g_i(x_i)- p_ig_\psi(\psi))$ represents the rate at which the $i$th substrate receives electrons from the $(i-1)$th substrate. Denoting by $f_i^{'}$, $g_i^{'}$ and $g_\psi^{'}$ the derivatives of the functions $f_i$, $g_i$ and $g_\psi^{'}$, the Jacobian of this system can be written $J = {J_0D}$ where ${J_0}$ is the symmetric matrix

\begin{equation}
J_0 = \left[
\begin{array}{ccccc}
-(f_1^{'}+f_2^{'}) & f_2^{'} & \cdots & 0 & p_1f_1^{'}-p_2f_2^{'}\\
f_2^{'} & -(f_2^{'}+f_3^{'}) & \cdots & 0 & p_2f_2^{'}-p_3f_3^{'}\\
\vdots & \vdots & \ddots & \vdots & \vdots \\
0 & 0 & \cdots & -(f_n^{'}+f_{n+1}^{'}) & p_nf_n^{'}-p_{n+1}f_{n+1}^{'}\\
p_1f_1^{'}\!-\!p_2f_2^{'} & p_2f_2^{'}\!-\!p_3f_3^{'} & \cdots & p_nf_n^{'}\!-\!p_{n+1}f_{n+1}^{'} & -\!\!\sum\limits_{i=1}^{n+1}p_i^2f_i^{'} - \frac{L_\psi}{g_\psi^{'}}
\end{array}
\right]
\end{equation}

and ${D}$ is the positive diagonal matrix

\begin{equation}
D = \left[
\begin{array}{ccccc}
g_1^{'} & 0 & \cdots & 0 & 0\\
0 & g_2^{'} & \cdots & 0 & 0\\
\vdots & \vdots & \ddots & \vdots & \vdots \\
0 & 0 & \cdots & g_{n}^{'} & 0\\
0 & 0 & \cdots & 0 & g_\psi^{'}
\end{array}
\right]
\end{equation}

From the discussions earlier, ${J_0}$ is a $P^{(-)}$ matrix. Further it is symmetric, and hence sign symmetric (see Appendix~\ref{appdefs} for a definition of sign symmetry). This implies \cite{kafri} that ${J_0}$ is $D$-stable, i.e. the product of ${J_0}$ with any positive diagonal matrix is Hurwitz. Hence $J$ is Hurwitz. Thus the assumption that reaction rates depend on the sum of potentials of the half reactions involved ensures that the Jacobian of the system is everywhere Hurwitz.

\section{Discussion and conclusions}

We have analysed in some detail, and using a variety of mathematical techniques, the behaviour of electron transport chains coupled to a charge translocation process. In all cases trajectories are bounded, and there is a unique equilibrium, but questions about the stability of this equilibrium have proved harder. Where the chain consists of a single redox pair, the unique equilibrium is globally stable. When there are two redox pairs the same conclusions can be reached subject to some extra conditions on the feedback process. In higher dimensions no such general conditions could easily be found. Thus the length of the electron transport chain is crucial in deciding on stability of the equilibrium.

It is somewhat surprising that the coupling of electron transfer to a membrane potential -- a negative feedback loop -- can serve to destabilise the unique equilibrium in these systems. Interestingly, when the reaction rates are monotonic functions of a sum of potentials, then the system in any dimension could be proved to be everywhere Hurwitz. Reaction rates cannot in general be seen in this way, but in the case of reactions which are primarily about charge transfer, the assumption could be reasonable. Certainly some of the choices of reaction rates in numerical models such as \cite{korzeniewski2} satisfy this assumption.

There are some interesting open questions, both biological and mathematical. From a biological point of view, it is of interest to find out whether experiments on mitochondria with constant inputs ever display behaviour other  than convergence to an equilibrium, such as periodic or chaotic behaviour. If this is never the case, then this suggests that our very general model may be omitting certain important biological/thermodynamic restrictions on the reaction rates, which would tend to stabilise the system. It would also be interesting to see how additional processes such as transport processes in the full numerical models (\cite{korzeniewski2}, \cite{beard} for example) affect the conclusions presented here.

An open mathematical question is whether there are equivalent conditions to the ordering condition in 3D which ensure that the Jacobian of the system is Hurwitz in arbitrary dimension, or better still that the second additive compound has negative logarithmic norm, and hence the unique equilibrium is globally stable. If such conditions exist can they be given general biological meanings?

It would also be interesting to explore when the results presented here survive weakening of the assumption that electrons are transferred along a chain. Although electron transfers taking place in the mitochondrial membrane are often described via a ``chain'' it is likely that this description is a convenient simplification rather than the whole truth. General electron transfer networks in the absence of a potential were analysed in \cite{banajimathchem} and found to have simple behaviour. Application of the theory presented in \cite{banajiSIAM} should allow determination of when these networks give rise to $P^{(-)}$ Jacobians when interacting with a membrane potential.

Finally, although conditions ensuring sign-symmetry of the system imply that the Jacobian is everywhere Hurwitz, it is an open question as to whether this implies global stability of the unique equilibrium. Since the Markus-Yamabe conjecture does not hold in dimensions greater than 2 \cite{cima}, global stability does not follow automatically from local stability, and requires independent proof.

\appendix
\section{Definitions}
\label{appdefs}

\subsection{Hurwitz stability of matrices}

A square matrix is defined to be {\bf Hurwitz stable} if all its eigenvalues lie in the open left half of the complex plane -- i.e. the real parts of all its eigenvalues are negative.

\subsection{$P$ matrices and related classes}

For some $n \times m$ matrix $A$, $A(\alpha|\gamma)$ will refer to the submatrix of $A$ with rows indexed by the set $\alpha \subset \{1, \ldots, n\}$ and columns indexed by the set $\gamma \subset \{1, \ldots, m\}$. A {\bf principal submatrix} of $A$ is a submatrix containing columns and rows from the same index set, i.e. of the form $A(\alpha|\alpha)$. A {\bf minor} is the determinant of any square submatrix of $A$. If $A(\alpha|\gamma)$ is a square submatrix of $A$ (i.e. $|\alpha| = |\gamma|$), then $A[\alpha|\gamma]$ will refer to the corresponding minor, i.e. $A[\alpha|\gamma]= \mathrm{det}(A(\alpha|\gamma))$. A {\bf principal minor} of $A$ is the determinant of a principal submatrix of $A$.

$P$ matrices are square matrices all of whose principal minors are positive. They are by definition nonsingular. If $\,-A$ is a $P$ matrix, then we will say that $A$ is a $P^{(-)}$ matrix. If $A$ is a $P^{(-)}$ matrix, this means that each $k \times k$ principal minor of $A$ has sign $(-1)^k$.

\subsection{Sign symmetry}

An $n \times n$ matrix is {\bf sign-symmetric} if symmetrically placed minors have the same sign, i.e. $A[\alpha|\gamma]A[\gamma|\alpha] \geq 0$ for every $\alpha, \gamma \subset \{1, \ldots, n\}$ with $|\alpha| = |\gamma|$.

\subsection{Second additive compound matrices}

A brief definition of the second additive compound of any square matrix can be found in \cite{li_muldowney_2000}. For a more detailed discussion see \cite{allen}. For a 3D matrix
\begin{equation}
A = \left(\begin{array}{ccc}
a_{11} & a_{12} & a_{13} \\
a_{21} & a_{22} & a_{23} \\
a_{31} & a_{32} & a_{33}
\end{array}\right)
\label{A_matrix_definition}
\end{equation}
the second additive compound takes the form\footnote{In general, the second additive compound of a matrix $A$ has dimension $^dC_2$ where \(d = \mathrm{dim}(A)\). When  $\mathrm{dim}(A) = 3$, we get $\mathrm{dim}(A^{[2]}) = 3$ also, but this is not generally the case.}
\[A^{[2]} = \left(\begin{array}{ccc}
a_{11} + a_{22} & a_{23} & -a_{13} \\
a_{32} & a_{11} + a_{33} & a_{12} \\
-a_{31} & a_{21} & a_{22} + a_{33}
\end{array}\right)\]
This second additive compound was constructed using the standard lexicographic ordering of basis vectors. It is possible to construct a second additive compound using a different ordering, but such choices make no difference to the logarithmic norms of the matrix.

\subsection{Logarithmic norms}

If $\|\cdot\|$ denotes a vector norm on $\mathbb{R}^n$, and also the induced matrix norm on $n \times n$ matrices, then the logarithmic norm \cite{strom}, also known as a Lozinski\u{\i} measure, of an $n \times n$ matrix $A$ is defined by
\begin{equation}
\label{normeqn}
\mu(A) = \lim_{h \to 0^+}\frac{\|I + hA\|-1}{h}
\end{equation}

\section{Local stability in 3D}
\label{local3D}
In this appendix we prove local stability of the equilibrium in three dimensions, subject to the assumption in (\ref{hurwitz_sign_assumption}), using the Routh-Hurwitz theorem. Consider the characteristic polynomial of a matrix $A$:
\begin{equation}
|\lambda I - A| = \lambda^n + b_1\lambda^{n-1} + \ldots + b_{n-1}\lambda + b_n
\end{equation}
In this equation, $I$ is the $n \times n$ identity matrix, and the coefficients $b_i$ are the sums of all principal minors of $-A$ of dimension $i$. For a $P^{(-)}$ matrix, $b_i > 0$ for all $i$. Now define $b_k \equiv 0$ for all $k > n$, and construct a set of numbers $\Delta_i$ as follows:
\begin{equation}
\Delta_i = \left|\begin{array}{cccccccc}
b_1 &  1  &  0  &  0  &  0  & 0 & \cdots & 0 \\
b_3 & b_2 & b_1 &  1  &  0  & 0 & \cdots & 0 \\
b_5 & b_4 & b_3 & b_2 & b_1 & 1 & \cdots & 0 \\
\vdots & \vdots & \vdots & \vdots & \vdots & \ddots & \vdots & 0 \\
b_{2i-1} & b_{2i-2} & b_{2i-3} & b_{2i-4} & b_{2i-5} & b_{2i-6} & \cdots & b_i
\end{array}\right|
\end{equation}

The Routh-Hurwitz theorem states that $A$ is Hurwitz if and only if $\Delta_i > 0$ for all $i\leq n$. In three dimensions, we need to check that the three quantities
\begin{eqnarray}
\Delta_1 & = & b_1 \\
\Delta_2 & = & b_1 b_2 - b_3 \\
\Delta_3 & = & b_3(b_1 b_2 - b_3) = b_3 \Delta_2
\end{eqnarray}
are all positive. Since all the $b_i$ are positive, all three quantities are positive if and only if $\Delta_2>0$. This in turn follows (condition 12 in \cite{kafri}) if
\[
0 < a_{12}a_{23}a_{31}+a_{21}a_{32}a_{13} - 2a_{11}a_{22}a_{33}
\]
where \(a_{ij}\) are elements of \(A\). Substituting \(a_{ij}\) for the elements of the Jacobian and expanding using the open source symbolic algebra program Maxima \cite{maxima} gives the following condition:

\begin{eqnarray*}
a_{12}a_{23}a_{31}+a_{21}a_{32}a_{13} - 2a_{11}a_{22}a_{33} & = & F_{21}\,F_{32} \left( 2 p_3F_{3\psi} +2 p_1F_{1\psi} -p_3F_{1\psi} \right) \\
&& + f_{11}\,f_{22} \left( 2 p_3F_{3\psi} +2 p_1F_{1\psi} -p_1F_{3\psi} \right)\\
&& + \mbox{ positive terms}
\end{eqnarray*}

With the ordering assumption (\ref{hurwitz_sign_assumption}), we get:
\begin{eqnarray}
2 p_3F_{3\psi} +2 p_1F_{1\psi} -p_3F_{1\psi} &\ \geq\ & 0 \label{ordering_condition_1} \\
2 p_3F_{3\psi} +2 p_1F_{1\psi} -p_1F_{3\psi} &\ \geq\ & 0 \label{ordering_condition_2}
\end{eqnarray}

Thus the Jacobian is everywhere Hurwitz and hence the unique equilibrium of the system must be locally asymptotically stable. Note that the restriction (\ref{hurwitz_sign_assumption}) is stronger than necessary to ensure that \(J\) is Hurwitz, but no other set of conditions with a clear physical meaning that make the Jacobian Hurwitz have been discovered. Finding a set of necessary and sufficient conditions for \(J\) to be Hurwitz is a difficult problem.

\bibliographystyle{elsart-num}

\end{document}